\pdfoutput=1
\documentclass[conference]{IEEEtran}
\IEEEoverridecommandlockouts
\usepackage{cite}
\usepackage{amsmath,amssymb,amsfonts}
\usepackage{algorithmic}
\usepackage{graphicx}
\usepackage{textcomp}
\usepackage{xcolor}
\usepackage{booktabs}
\usepackage{multirow}
\usepackage{listings}
\usepackage[most]{tcolorbox}
\lstdefinelanguage{diff}{
  morecomment=[f][\color{red!65!black}]{-},
  morecomment=[f][\color{green!45!black}]{+},
  morecomment=[f][\color{blue!60!black}]{@@},
}
\lstdefinestyle{patch}{
  language=diff,
  basicstyle=\ttfamily\scriptsize,
  breaklines=true,
  columns=fullflexible,
  keepspaces=true,
  showstringspaces=false,
  aboveskip=2pt, belowskip=2pt,
}
\def\BibTeX{{\rm B\kern-.05em{\sc i\kern-.025em b}\kern-.08em
    T\kern-.1667em\lower.7ex\hbox{E}\kern-.125emX}}
\begin{document}

\title{DepRepair: LLM-Based Source-Code Repair for Dependency Breaking Changes}

\author{
\IEEEauthorblockN{
Shenghao Yang\IEEEauthorrefmark{1},
Bo Lu\IEEEauthorrefmark{2},
Yaochen Liu\IEEEauthorrefmark{2},
Yu Kang\IEEEauthorrefmark{2},
Qiongfang Zhang\IEEEauthorrefmark{2},
Chetan Bansal\IEEEauthorrefmark{2},
Saravan Rajmohan\IEEEauthorrefmark{2},
Minghua Ma\IEEEauthorrefmark{2}
}
\IEEEauthorblockA{\IEEEauthorrefmark{1}Carnegie Mellon University, Pittsburgh, PA, USA\\
Email: shenghay@andrew.cmu.edu}
\IEEEauthorblockA{\IEEEauthorrefmark{2}Microsoft\\
Email: minghuama@microsoft.com}
\thanks{Work done during an internship at Microsoft.}
}

\maketitle

\begin{abstract}
Modern software projects depend on numerous third-party libraries, whose updates often introduce breaking changes. Adapting consumer code to such changes remains labor-intensive and error-prone. Existing work either characterizes dependency breaking changes without producing a verified consumer-side patch, or studies automated repair only in settings where the failure and repair context are contained within the target repository. However, dependency breaking changes violate this assumption: the decisive repair evidence lies upstream in release notes and API diffs, and no failing test localizes where the consumer breaks, leaving the repair under-informed. To study this cross-repository problem on real data, we introduce DepBench, a benchmark of 95 real-world dependency-update instances across four ecosystems, each paired with a Docker-based executable oracle that runs the consumer's own tests. To address these challenges, we propose \textsc{DepRepair}, a single-call LLM approach that grounds repair in structured upstream evidence through three components: an evidence filter that distills relevant upstream changes, a usage locator that identifies affected consumer sites, and a subcategory-aware guide that tailors repairs to the breaking-change type. Evaluated on \textsc{DepBench}, \textsc{DepRepair} attains the highest executable pass rate on each backbone, achieving 89.5\% with GPT-5.5 and 82.1\% with Claude Opus 4.6. We further find that raw upstream evidence reduces LLM and agent pass rates by 7--23 percentage points, whereas structured evidence consistently improves them.
\end{abstract}


\section{Introduction}\label{sec:intro}

Software projects depend heavily on third-party libraries, and a modern application typically pulls in dozens of direct and transitive dependencies that it must keep up to date for security fixes, new features, and performance improvements. When upstream libraries update, they often introduce breaking changes, such as API migrations, renames, import migrations, and paradigm shifts~\cite{raemaekers2017semantic}. Such changes usually cause compilation errors or runtime failures in downstream consumer code, forcing developers to manually adapt their projects to the new library version.

However, adapting consumer code to such changes is time-consuming and error-prone. Studies show that developers frequently delay dependency updates primarily because of the perceived cost~\cite{bogart2016break}, and 81.5\% of studied projects keep outdated dependencies~\cite{kula2018developers}. The resulting technical lag exposes projects to known vulnerabilities and missed performance gains~\cite{lu2025minimizing}. Thus, the tension between the cost of manual adaptation and the risk of falling behind motivates automated tools that repair the consumer code so that projects can keep their dependencies up to date.

Research on dependency management has emerged to understand and tame breaking changes from dependency updates.
One category \emph{characterizes} breaking changes, measuring how often upstream updates break their consumers and how the resulting risk propagates through the software supply chain~\cite{raemaekers2017semantic,decan2019empirical,zimmermann2019small}.
The other category builds tooling that \emph{assists} migration by reusing how the library or its earlier clients evolved~\cite{henkel2005catchup,teyton2012mining}.
However, for a specific project, most of this work neither localizes the exact consumer sites that must change nor emits a verified source patch that adapts the project to the new version.
As a result, a developer facing a concrete upgrade is still left to find and edit every affected line by hand.

In contrast, rule-based and LLM-based methods focus on localizing and repairing the broken consumer code.
Rule-based approaches rewrite outdated API usages into their replacements using transformation rules that are hand-crafted or mined from past migrations~\cite{dig2006apis,henkel2005catchup}.
LLM-based program repair instead synthesizes a patch directly from the consumer code and a failing test, assuming the fix can be found within the project itself.
This paradigm was established by SWE-bench~\cite{jimenez2024swe} and now drives coding agents such as Codex~CLI~\cite{openai2025codex} and Claude~Code~\cite{anthropic2025claudecode}.
Despite their difference in form, both treat dependency breaking-change repair as a same-repository problem, overlooking its inherently cross-repository nature.
This leaves the repair under-informed and the model prone to abstain or mis-apply the fix, since the evidence for what to change resides upstream in release notes and API diffs, and no failing test localizes where the breakage occurs.
For example, when a project upgrades JUnit from~4 to~5, the required \texttt{@Before}$\to$\texttt{@BeforeEach} rename appears only in the JUnit~5 release notes, not in the consumer's own code or tests.

To study this cross-repository problem in real data, we need a benchmark that no prior work provides: its instances must be driven by real dependency upgrades, carry the upstream evidence required to repair them, and be verified by running the consumer's own tests.
We therefore design a four-stage construction strategy to establish such a benchmark. It first mines dependency-update pull requests from GitHub, then filters them for repository and pull-request quality, next confirms the changes are caused by the upgrade and backed by upstream evidence, and finally keeps only instances whose upgrade breaks the consumer's tests and whose patch repairs them in Docker.
Through this strategy, we construct \textsc{DepBench}, which includes 95 real-world instances across four ecosystems---Maven~(38), npm~(33), Cargo~(15), and PyPI~(9)---each paired with a Docker-based executable oracle that runs the consumer's own test suite.

Constructing \textsc{DepBench} and analyzing its instances let us pinpoint the central challenges of cross-repository breaking-change repair.
First, the decisive repair evidence is external and noisy, so the correct fix must be distilled from upstream release notes rather than read off the consumer code. In 57 of the 95 instances (60\%) the developer's patch adopts an upstream API symbol (e.g., a new import or namespace) that never appears in the consumer's pre-upgrade sources.
Second, localization gets no help from a failing test, so the affected call sites and imports must be found by other means and are widely diffuse. In \textsc{DepBench}, the gold patches span a median of 10 files and 21 edit hunks, and 85 of the 95 (89\%) touch more than one file.
Third, no single repair strategy fits all, so the edit must be tailored to the breaking-change type, which ranges across direct renames~(27), compound API migrations~(25), import migrations~(23), and paradigm shifts~(20).

To address these challenges, we propose \textsc{DepRepair}, a single-call LLM approach that treats dependency repair as an evidence-grounding problem.
Specifically, \textsc{DepRepair} structures the upstream evidence with three key components: (1)~\emph{upstream evidence filter}, which first collects and distills breaking-change information from the upstream library into a compact set of migration rules; (2)~\emph{consumer-side usage locator}, which then identifies the specific call sites and imports affected in the downstream project; and (3)~\emph{subcategory-aware guide}, which finally tailors the repair instructions to the type of breaking change (e.g., API migration, rename, import migration, or paradigm shift).

To systematically evaluate \textsc{DepRepair} and isolate the role of cross-repository evidence, we investigate four research questions (RQs):
\begin{itemize}
\item \textbf{RQ1 (Overall comparison):} How does \textsc{DepRepair} compare with prompt baselines and state-of-the-art coding agents, and are the differences statistically significant? This establishes whether grounding a single LLM call in structured upstream evidence can match multi-turn agents that repeatedly read and edit the repository.
\item \textbf{RQ2 (Failure taxonomy):} When repair fails, \emph{how} does it fail? Characterizing the failure modes reveals whether the bottleneck is producing \emph{any} fix or producing a \emph{correct} one, which dictates where evidence should intervene.
\item \textbf{RQ3 (Evidence form and ablation):} How does the \emph{form} of upstream evidence---and each \textsc{DepRepair} component---affect repair for both LLMs and agents? This tests our central hypothesis that the value of cross-repository context lies in its \emph{processing}, not its mere \emph{provision}.
\item \textbf{RQ4 (Subcategory analysis):} How does performance vary across breaking-change subcategories, and where does structured evidence help most?
\end{itemize}

We summarize our key findings as follows.
For \textbf{RQ1}, \textsc{DepRepair} attains the highest executable pass rate on each backbone---89.5\% with GPT-5.5 and 82.1\% with Claude~Opus~4.6---matching or surpassing multi-turn coding agents with a single LLM call, which shows that the bottleneck is access to the right cross-repository evidence rather than the amount of in-repository interaction.
For \textbf{RQ2}, the empty patch dominates: the most common failure is emitting \emph{no} usable patch rather than an incorrect one, and raw upstream evidence aggravates it---Direct LLM (upstream) returns an empty patch on 32 of 95 instances, whereas \textsc{DepRepair} cuts this to 10.
For \textbf{RQ3}, cross-repository evidence helps only when structured---supplying the raw upstream changelog lowers pass rates by up to 10.5 percentage points for LLMs and up to 23.2 for agents, whereas the same evidence distilled into migration rules consistently improves them.
For \textbf{RQ4}, \textsc{DepRepair}'s gains concentrate on direct-rename and compound API migrations, whereas paradigm shifts---which demand structural refactoring beyond rule-level edits---remain the hardest residual, solved on only 65--70\% of cases even by the best method.

Our contributions are:
\begin{itemize}
\item We shift the perspective on dependency breaking-change repair from a same-repository task to a cross-repository one, recognizing that the decisive repair evidence lies upstream rather than in the consumer project.
\item We introduce \textsc{DepBench}, a benchmark of 95 real-world dependency-update instances across four ecosystems, each with a verified ground-truth patch and a Docker-based executable oracle.
\item We propose \textsc{DepRepair}, a single-call approach that grounds the model in distilled upstream evidence and tailors its guidance to the breaking-change type, matching or surpassing state-of-the-art coding agents at a fraction of their cost.
\item We show that raw upstream evidence hurts both LLMs and agents while distilled evidence consistently helps, and that the \emph{empty patch}---not incorrect repair---is the dominant failure mode, which structured evidence sharply reduces (from 32 to 10 of 95 instances).
\end{itemize}

\section{Related Work}\label{sec:related}

\subsection{Dependency Updates and Breaking Changes}
Prior work on dependency management measures the breaking-change problem or assists migration, but none automatically generates the consumer-side source patch that adapts a downstream project to an upstream update.
One line \emph{characterizes} the problem empirically, quantifying the prevalence and client-side impact of breaking changes, behavioral incompatibilities, and semantic-versioning violations across ecosystems~\cite{raemaekers2017semantic,xavier2017historical,brito2018and,ochoa2022breaking,decan2019empirical,mostafa2017experience,jayasuriya2023understanding}, showing that most projects run outdated dependencies and negotiate breaking changes and API deprecations at a real cost~\cite{kula2018developers,bogart2016break,dig2006apis,derr2017keep,sawant2019react}.
This breaking-change burden is what keeps projects from staying current, with direct consequences for the software supply chain: stale and bloated dependencies, as well as dependency conflicts, expose projects to known vulnerabilities and supply-chain attacks~\cite{decan2018impact,pashchenko2018vulnerable,zimmermann2019small,ohm2020backstabber,ladisa2023sok,soto2021comprehensive,wang2018dependency,wang2020watchman,wang2020empirical,liu2022demystifying}, yet even automated upgrade pull requests are frequently ignored because they break the consumer's build~\cite{cox2015measuring,mirhosseini2017can,alfadel2021use}.
A second line provides \emph{tooling} that supports, but does not fully automate, migration: Henkel and Diwan~\cite{henkel2005catchup} record and replay library-side API refactorings (CatchUp!), Lu et al.~\cite{lu2025minimizing} propose strategies for minimizing breaking changes when mitigating technical lag, and earlier studies chart how libraries are migrated in practice~\cite{cossette2012seeking,teyton2012mining} and test for breaking changes across versions~\cite{mezzetti2018type}.
\textsc{DepRepair} closes this gap: rather than measuring or recording breaking changes, it distills upstream evidence into structured migration rules and uses them to generate the consumer-side source patch that resolves the breakage.

\subsection{Same-Repository Repair}
Existing automated repair for breaking changes, whether rule-based or LLM-based, shares a common assumption: the information needed to fix the code already lives \emph{inside} the consumer project (or in rules derived from past migrations of it).
We organize prior work along these two lines and contrast both with our cross-repository view.

\subsubsection{Rule-Based Repair}
Rule-based approaches rewrite outdated API usages into their replacements using transformation rules.
Dig and Johnson characterized how APIs evolve through refactorings~\cite{dig2006apis}, and Henkel and Diwan's CatchUp! records refactorings in the library and replays them on consumers~\cite{henkel2005catchup}.
To avoid hand-written rules, later work mines adaptations automatically: Dagenais and Robillard's SemDiff recommends replacement methods by analyzing how the framework itself adapts to its own changes~\cite{dagenais2011recommending}, Nguyen et al.'s LibSync learns API-usage adaptation patterns from already-migrated clients~\cite{nguyen2010graph}, Fazzini et al.'s AppEvolve infers and applies update edits from examples of other projects that completed the same migration~\cite{fazzini2019automated}, Xu et al.'s Meditor infers and applies API migration edits from prior client commits~\cite{xu2019meditor}, and Lamothe et al.'s A3 assists API migrations using code examples~\cite{lamothe2020a3}.
These methods pre-encode the fix as transformation rules; they depend on either curated rules or a corpus of prior migrations, and they struggle when the upstream evidence is unstructured release notes rather than clean before/after examples.
\textsc{DepRepair} instead distills such raw upstream evidence into migration rules at repair time, removing the dependence on pre-existing rule sets or migrated exemplars.

\subsubsection{LLM-Based Repair}
Automated program repair has a long history, and large language models now drive its strongest results~\cite{xia2023automated,fan2023automated,jin2023inferfix,hou2024large}.
SWE-bench established real-world GitHub issues as a benchmark for this setting~\cite{jimenez2024swe}, spurring a range of repair systems.
Agentless follows a non-agentic, two-phase localize-then-repair pipeline~\cite{xia2024agentless}, whereas SWE-agent~\cite{yang2024swe}, OpenHands~\cite{wang2025openhands}, AutoCodeRover~\cite{zhang2024autocoderover}, RepairAgent~\cite{bouzenia2025repairagent}, and repository-level planners such as CodePlan~\cite{bairi2024codeplan} grant the model autonomy to navigate and edit repository files over multiple turns, and commercial agents such as Codex~CLI~\cite{openai2025codex} and Claude~Code~\cite{anthropic2025claudecode} apply the same paradigm in practice.
All of these assume a \emph{same-repository} setting: the bug report, relevant code, and fix reside within a single project, and the model recovers the fix by reading and editing that project.
Dependency breaking-change repair violates this assumption, because the decisive evidence lies in the upstream library; \textsc{DepRepair} addresses this by explicitly grounding a single LLM call in distilled cross-repository evidence.

\section{Benchmark}\label{sec:benchmark}

\subsection{Construction}\label{sec:construction}
\begin{figure*}[htbp]
\centerline{\includegraphics[width=\textwidth]{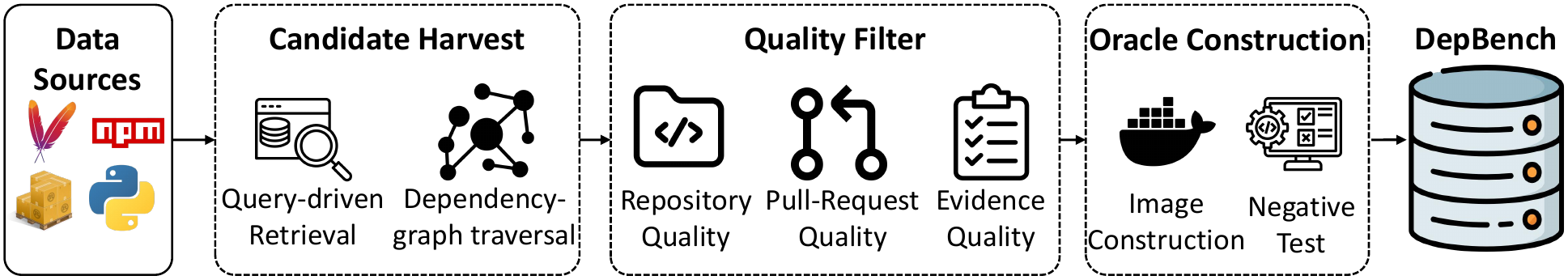}}
\caption{Construction pipeline for \textsc{DepBench}. Candidate pull requests are first harvested from popular packages across four ecosystems by query-driven retrieval and dependency-graph traversal. They are then passed through quality filters that retain only genuine, reproducible, and evidence-bearing PRs, and finally paired with a Docker-based executable oracle, yielding 95 self-contained instances.}
\label{fig:benchmark}
\end{figure*}

We construct \textsc{DepBench}, a benchmark of real-world dependency update instances drawn from open-source projects.
Each instance represents a pull request that updates a dependency to a new major version and modifies consumer source code to adapt to breaking API changes introduced by the update.
Fig.~\ref{fig:benchmark} summarizes the construction pipeline, which proceeds from candidate harvesting through quality filtering to oracle construction.

\paragraph{Data sources.}
We draw from four popular package ecosystems: Maven, npm, Cargo, and PyPI, covering both statically and dynamically typed languages.
Within them we target 10 popular libraries undergoing a well-documented major-version transition: JUnit~4$\to$5 and Spring~Boot~2.7$\to$3.0 (Maven), Vue~2$\to$3, Next.js~12$\to$13, and ESLint~8$\to$9 (npm), Clap~3$\to$4 and Tokio~0.3$\to$1.0 (Cargo), and Pydantic~1.10$\to$2.0, SQLAlchemy~1.4$\to$2.0, and NumPy~1.26$\to$2.0 (PyPI).

\paragraph{Candidate harvest.}
To collect genuine migration pull requests with high recall, we combine two complementary strategies.
\emph{Query-driven retrieval} uses keywords to search GitHub for pull requests that both bump a dependency version in a manifest file (e.g. \texttt{pom.xml}, \texttt{package.json}, \texttt{Cargo.toml}, or \texttt{requirements.txt}) and modify source files in the same commit.
\emph{Dependency-graph traversal} follows each library's reverse dependency graph to consumers still pinning the old major version, surfacing migration commits that keyword search misses.

\paragraph{Quality Filter.}
To retain only instances that are genuine, reproducible, and evidence-bearing, each candidate must clear three gates.
\emph{Repository quality}: the project must build, excluding forks and abandoned repositories, so an executable environment is reconstructable.
\emph{Pull-request quality}: the source edits must be caused by the update, excluding rollbacks, lockfile- or manifest-only changes, unrelated build churn, and patches touching files outside the consumer snapshot.
\emph{Evidence quality}: usable upstream signals---release notes, changelogs, or an API diff---must exist, since these are the inputs that \textsc{DepRepair} and the upstream baselines consume.

\paragraph{Oracle construction.}
We package the pre-update snapshot, update metadata, upstream evidence, and the developer's gold patch into a self-contained Docker image, then probe it with two checks: a \emph{negative test} confirming the un-repaired snapshot actually fails the breakage-related tests, and a \emph{positive test} confirming the gold patch makes those tests pass.
Building successfully is necessary but not sufficient: we keep only \emph{feasible} instances---those exhibiting real, reproducible breakage that the gold patch demonstrably repairs.
The 95 feasible instances surviving all three stages constitute \textsc{DepBench}.

\subsection{Breaking-Change Taxonomy}\label{sec:taxonomy}
We classify each instance into one of four subcategories based on the type of upstream breaking change:

\begin{itemize}
\item \textbf{Direct rename} (27 instances): Symbols, types, or import paths are renamed in the upstream API (e.g., \texttt{clap::App} $\to$ \texttt{clap::Command}).
\item \textbf{Compound API migration} (25): Multiple related APIs change simultaneously, requiring coordinated multi-site modifications (e.g., Spring Boot annotation and property changes).
\item \textbf{Import migration} (23): The module structure is reorganized upstream, requiring import path updates throughout the consumer (e.g., JUnit~4 $\to$ JUnit~5 namespace migration).
\item \textbf{Paradigm shift} (20): The upstream API's design philosophy changes fundamentally (e.g., Vue~2 Options API $\to$ Vue~3 Composition API), requiring structural refactoring beyond simple renaming.
\end{itemize}

\subsection{Benchmark Statistics}\label{sec:stats}
The final benchmark comprises 95 instances across four package ecosystems, partitioned cleanly into the four breaking-change subcategories above, as shown in Table~\ref{tab:benchmark}.

\begin{table}[htbp]
\caption{Benchmark distribution by ecosystem and breaking-change subcategory.}
\begin{center}
\begin{tabular}{lcccc|c}
\toprule
\textbf{Subcategory} & \textbf{Maven} & \textbf{npm} & \textbf{Cargo} & \textbf{PyPI} & \textbf{Total} \\
\midrule
Direct rename         & 0  & 6  & 15 & 6 & 27 \\
Compound API migr.    & 19 & 4  & 0  & 2 & 25 \\
Import migration      & 19 & 4  & 0  & 0 & 23 \\
Paradigm shift        & 0  & 19 & 0  & 1 & 20 \\
\midrule
\textbf{Total}        & \textbf{38} & \textbf{33} & \textbf{15} & \textbf{9}  & \textbf{95} \\
\bottomrule
\end{tabular}
\label{tab:benchmark}
\end{center}
\end{table}

Table~\ref{tab:packages} shows the target packages and their version transitions.
Each instance provides on the order of ten source files as context to the tool, and the ground-truth patches modify tens of diff hunks on average (up to several hundred), reflecting the non-trivial nature of real-world migration edits.

\begin{table}[htbp]
\caption{Target packages and version transitions.}
\begin{center}
\begin{tabular}{llrl}
\toprule
\textbf{Package} & \textbf{Ecosystem} & \textbf{$n$} & \textbf{Version} \\
\midrule
Vue          & npm      & 20 & 2 $\to$ 3 \\
JUnit        & Maven    & 19 & 4 $\to$ 5 \\
Spring Boot  & Maven    & 19 & 2.7 $\to$ 3.0 \\
Clap         & Cargo    & 11 & 3 $\to$ 4 \\
Next.js      & npm      & 8  & 12 $\to$ 13 \\
Pydantic     & PyPI     & 6  & 1.10 $\to$ 2.0 \\
ESLint       & npm      & 5  & 8 $\to$ 9 \\
Tokio        & Cargo    & 4  & 0.3 $\to$ 1.0 \\
SQLAlchemy   & PyPI     & 2  & 1.4 $\to$ 2.0 \\
NumPy        & PyPI     & 1  & 1.26 $\to$ 2.0 \\
\bottomrule
\end{tabular}
\label{tab:packages}
\end{center}
\end{table}

\section{Approach}\label{sec:approach}

\begin{figure*}[htbp]
\centerline{\includegraphics[width=\textwidth]{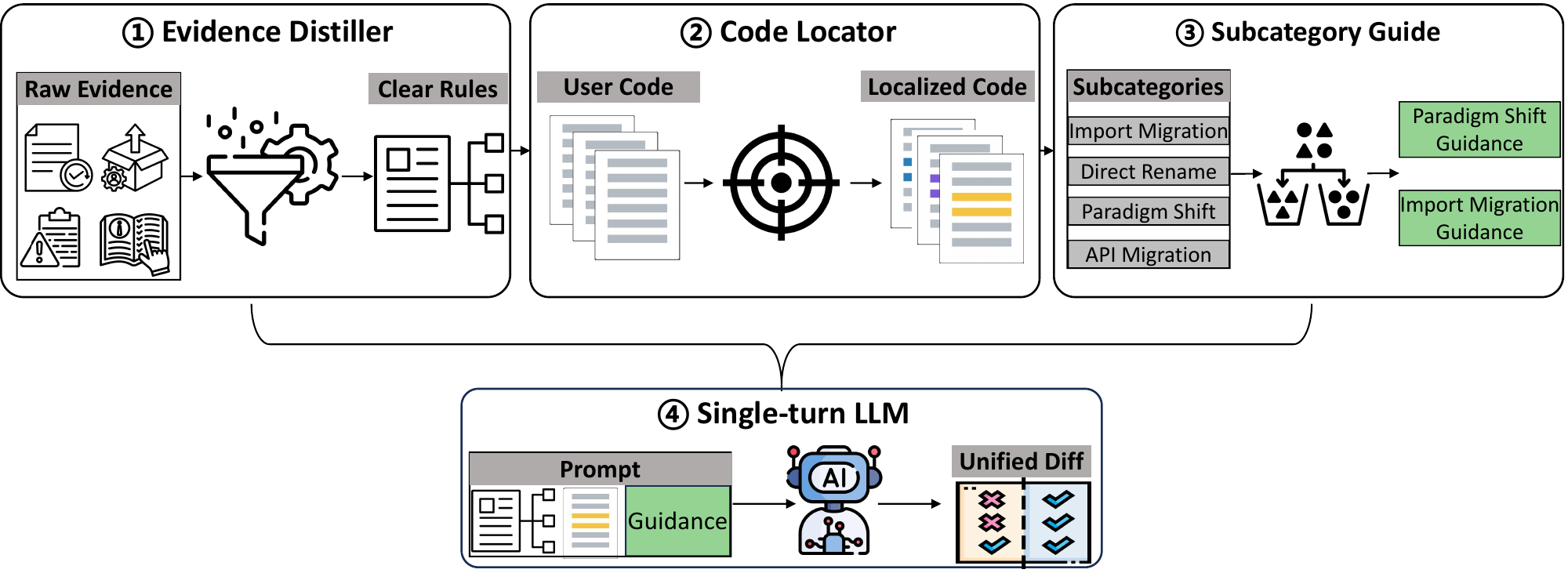}}
\caption{Overview of the \textsc{DepRepair} approach. Three evidence transformations: an \emph{evidence filter} that distills raw upstream changelogs into structured migration rules (\emph{what}), a \emph{usage locator} that connects those rules to affected consumer call sites (\emph{where}), and a \emph{subcategory guide} that tailors the repair strategy to the breaking-change type (\emph{how}). Finally, all these are assembled into a single LLM call that emits a unified diff, verified by the Docker-based executable oracle.}
\label{fig:overview}
\end{figure*}

\textsc{DepRepair} grounds the entire repair in the upstream evidence.
Because raw evidence is not directly actionable, \textsc{DepRepair} structures it into the three decisions a repair must make: \emph{what} changed upstream, \emph{where} it affects the consumer, and \emph{how} to edit. Each decision is produced by a dedicated transformation and composed into a single LLM call, as shown in Fig.~\ref{fig:overview}.
We use the JUnit~4$\to$5 migration \texttt{exec-114} as a running example throughout this section.

\subsection{Problem Formulation}\label{sec:formulation}
We formalize dependency breaking-change repair as a function of three inputs.
The \emph{dependency update} $u=(p, v_{\mathit{old}}, v_{\mathit{new}})$ names the package $p$ and its old and new versions.
The \emph{consumer snapshot} $C=\{f_1,\dots,f_n\}$ is the set of source files in the project before the update is applied.
The \emph{upstream evidence} $E$ is the unstructured documentation associated with $u$, such as release notes, changelogs, migration guides, and version diffs.
The goal is to produce a patch $\Delta$ (a unified diff over $C$) such that, after applying $\Delta$ and bumping $p$ to $v_{\mathit{new}}$, the consumer's own test suite passes.

This formulation makes the \emph{cross-repository} nature of the task explicit: the signal that determines a correct $\Delta$ is concentrated in $E$, which originates outside $C$, and unlike conventional repair there is no failing test inside $C$ that localizes the fault.
\textsc{DepRepair} produces $\Delta$ in a single generation step that composes three evidence-derived inputs:
\begin{equation}
\Delta = \mathrm{LLM}\big(\textsc{Assemble}(R, L, g, C)\big),
\label{eq:pipeline}
\end{equation}
where the migration rules $R$, the localization hint $L$, and the subcategory instruction $g$ are produced by three transformations of the inputs:
\begin{equation}
R = \textsc{Filter}(u, E), 
\label{eq:filter}
\end{equation}
\begin{equation}
L = \textsc{Localize}(C, R),
\label{eq:localize}
\end{equation}
\begin{equation}
g = \textsc{Guide}(u).
\label{eq:guide}
\end{equation}

The four subsections below develop these operators in turn.

\subsection{Evidence Filter: \emph{what} changed}\label{sec:extract}
To address the noise interference caused by raw upstream evidence, the evidence filter distills the raw evidence $E$ into a compact, structured set of \emph{migration rules} $R$.
Specifically, it first gathers candidate evidence for the version range $(v_{\mathit{old}}, v_{\mathit{new}})$ from sources that are routinely available across ecosystems:
\begin{itemize}
\item release notes and changelogs from the package registry (Maven Central, npm, crates.io, PyPI);
\item upstream source diffs between the two versions, when the project is open source;
\item deprecation and removal markers in API documentation; and
\item dedicated migration guides, when the maintainers publish them.
\end{itemize}

However, this evidence is often noisy, mixing a few consumer-visible breaking changes with many internal refactorings, performance notes, and unrelated features.
Thus, the filter then prompts the backbone LLM to distill this evidence into a list of rules, where each rule $r\in R$ records the \emph{old} API surface, its \emph{new} replacement, and a one-line description of the required edit, discarding the surrounding prose.
For our running example \texttt{exec-114}, this stage yields rules for the JUnit~4$\to$5 namespace move and annotation renames, such as:
\begin{quote}\small\ttfamily
\{old: "org.junit.Before", new: "org.junit.jupiter.api.BeforeEach", kind: import+annotation rename\}
\end{quote}
so that downstream reasoning operates on actionable facts rather than narrative text.

\subsection{Usage Locator: \emph{where} it matters}\label{sec:localize}
The second stage, $\textsc{Localize}$, identifies which parts of the consumer are affected by the rules in $R$, producing a \emph{localization hint} $L$.
Whereas conventional repair localizes from a failing test or a bug report, no such signal exists here; our localization is instead \emph{API-driven}, connecting each rule's old API surface to its usage sites in $C$.
For every rule $r\in R$ we scan the consumer files for:
\begin{itemize}
\item imports of changed or removed symbols;
\item call sites of modified APIs;
\item code patterns matching renamed or restructured constructs; and
\item transitive uses reached through local wrappers or re-exports.
\end{itemize}

The result is a list of (file, line, matched rule) anchors that narrows the model's attention to the regions that actually need editing.
On \texttt{exec-114}, $\textsc{Localize}$ flags the JUnit imports and the \texttt{@Before}/\texttt{@Test} annotation sites across the test files, pointing the model straight at the lines to change.

\subsection{Subcategory Guide: \emph{how} to edit}\label{sec:guide}
Knowing \emph{what} changed and \emph{where} it is used still leaves open \emph{how} to edit: the same rules call for a mechanical sweep in one case and a structural rewrite in another.
The third stage, $\textsc{Guide}$, supplies the model with an editing strategy matched to the kind of change at hand.
It first classifies the change type based on the migration rules $R$: pure symbol substitutions as a \emph{direct rename}, relocated import paths as an \emph{import migration}, several interacting rules as a \emph{compound API migration}, and rules that alter a design contract as a \emph{paradigm shift}. It then attaches the matching instruction $g$ that tells the model how to apply the rules:
\begin{itemize}
\item for \emph{direct renames}, $g$ emphasizes systematic, exhaustive symbol replacement across all occurrences;
\item for \emph{import migrations}, $g$ stresses updating import paths consistently while leaving call sites intact;
\item for \emph{compound API migrations}, $g$ warns that several related changes must be applied together or the patch will regress; and
\item for \emph{paradigm shifts}, $g$ instructs the model to perform structural refactoring guided by the migration rules rather than a one-to-one substitution.
\end{itemize}

On \texttt{exec-114}, the rules touch a namespace, several annotations, and an assertion API at once, so $\textsc{Guide}$ reads the change as a compound API migration and tells the model to apply the three in concert.

\subsection{Assembly and Single-Call Generation}\label{sec:assemble}
Finally, $\textsc{Assemble}$ composes the four ingredients into a single prompt: system instructions fixing the output format, the migration rules $R$, the localization hint $L$, the subcategory instruction $g$, and the consumer source $C$.
The backbone LLM answers in one call with a unified diff $\Delta$, which we apply to the consumer snapshot and verify with the Docker-based executable oracle (Section~\ref{sec:evaluation}).

\section{Evaluation}\label{sec:evaluation}

\subsection{Research Questions}
We investigate four research questions:
\begin{itemize}
\item \textbf{RQ1} (Overall comparison): How does \textsc{DepRepair} compare with baseline approaches and state-of-the-art coding agents, and are the differences statistically significant?
\item \textbf{RQ2} (Failure taxonomy): When repair fails, \emph{how} does it fail, and what does the distribution of failure modes reveal about the underlying difficulty?
\item \textbf{RQ3} (Evidence form and ablation): How does the \emph{form} of upstream evidence---and each \textsc{DepRepair} component---affect repair performance for both LLMs and agents?
\item \textbf{RQ4} (Subcategory analysis): How does repair performance vary across breaking-change subcategories, and where does structured evidence help most?
\end{itemize}

\subsection{Experimental Setup}
\textbf{Models.}
We evaluate with two LLM backbones to ensure generalizability: GPT-5.5 (OpenAI) and Claude Opus~4.6 (Anthropic), both set with a 300-second timeout and temperature~0.

\textbf{Direct LLM baselines.}
Our method \textsc{DepRepair} is a single-call (non-agentic) LLM approach. We therefore compare it against the same single-call interface without structured evidence, which we term \emph{Direct LLM}~\cite{xia2023automated}, prompting each backbone once to emit a unified diff under two evidence conditions:
\begin{itemize}
\item \textbf{\textsc{DepRepair}} (our\_method): Full pipeline with structured migration rules, localization hints, and subcategory-aware instructions; unified diff output.
\item \textbf{Direct LLM (upstream)}: Raw upstream changelog/migration guide included verbatim; unified diff output.
\item \textbf{Direct LLM (minimal)}: Only dependency update metadata and consumer code; no upstream evidence; unified diff output.
\end{itemize}

\textbf{Agentic baselines.}
We compare against two state-of-the-art coding agents that follow the agentic, multi-turn repair paradigm~\cite{yang2024swe}, each evaluated in both \emph{minimal} and \emph{upstream} configurations:
\begin{itemize}
\item \textbf{Codex~CLI}~\cite{openai2025codex}: OpenAI's coding agent (GPT-5.5 backbone) with multi-turn file editing.
\item \textbf{Claude~Code}~\cite{anthropic2025claudecode}: Anthropic's coding agent (Claude~Opus~4.6 backbone) with autonomous planning and editing.
\end{itemize}

\textbf{Controlled generation setting.}
To isolate the role of upstream evidence, every method---prompt baselines and agents alike---generates its patch from the same consumer snapshot without an in-loop build/test environment, and the resulting patch is verified afterwards by the shared Docker oracle.
This holds the execution interface fixed across methods, so that differences in pass rate reflect the \emph{form} of cross-repository evidence rather than the amount of in-repository execution feedback available during generation.

\textbf{Evaluation oracle.}
We use an \emph{execution-based oracle}, which grounds every verdict in the consumer's own tests rather than textual similarity to the developer patch.
For each instance, we build the consumer project in a Docker container, apply the candidate patch, and run the project's test suite.
An instance is counted as \textsc{Pass} if all breakage-related tests pass, and \textsc{Fail} otherwise.
Instances where the LLM times out or returns an API error are counted as failures.
We report the \emph{pass rate}: $\textsc{Pass}/95$ across all feasible instances.

\subsection{RQ1: Overall Comparison}\label{sec:rq1}

Table~\ref{tab:results} presents the main results across all methods and both backbones.

\begin{table}[htbp]
\caption{Overall comparison on 95 benchmark instances, grouped by backbone. Pass rate = \textsc{Pass}/95. On each backbone, \textsc{DepRepair} is significantly higher than every other method (McNemar paired test, $p<0.05$) except the runner-up ($\dagger$), whose gap to \textsc{DepRepair} is not statistically significant.}
\begin{center}
\begin{tabular}{lrc}
\toprule
\textbf{Method} & \textbf{Pass} & \textbf{Rate} \\
\midrule
\multicolumn{3}{l}{\emph{GPT-5.5 backbone}} \\
\textsc{DepRepair}     & \textbf{85} & \textbf{89.5\%} \\
Agentic Codex (minimal)  & 83 & 87.4\%$\dagger$ \\
Direct LLM (minimal)   & 78 & 82.1\% \\
Direct LLM (upstream)  & 71 & 74.7\% \\
Agentic Codex (upstream) & 70 & 73.7\% \\
\midrule
\multicolumn{3}{l}{\emph{Claude Opus~4.6 backbone}} \\
\textsc{DepRepair}     & \textbf{78} & \textbf{82.1\%} \\
Agentic Claude~Code (minimal)  & 77 & 81.1\%$\dagger$ \\
Direct LLM (minimal)   & 68 & 71.6\% \\
Direct LLM (upstream)  & 58 & 61.1\% \\
Agentic Claude~Code (upstream) & 55 & 57.9\% \\
\bottomrule
\end{tabular}
\label{tab:results}
\end{center}
\end{table}

We test each pairwise gap to \textsc{DepRepair} with the \emph{McNemar paired test} (exact binomial form when the discordant count is small).

\begin{itemize}
\item \textbf{\textsc{DepRepair} attains the highest pass rate.}
On both backbones it tops every baseline and agent, reaching 89.5\% on GPT-5.5 and 82.1\% on Opus~4.6.
\item \textbf{\textsc{DepRepair} significantly beats the same-backbone Direct LLM.}
\textsc{DepRepair} adds $+$7 instances on GPT-5.5 ($p{=}0.016$) and $+$10 on Opus~4.6 ($p{=}0.002$) over Direct LLM (minimal), with \emph{no} instance regressing in the other direction ($b_{01}{=}0$ on both backbones)---a clean, one-sided improvement.
\item \textbf{\textsc{DepRepair} matches, but does not beat, the strongest agent.}
The closest competitors are the cross-architecture agents in their minimal configurations---Codex (87.4\%) and Claude~Code (81.1\%). \textsc{DepRepair}'s edge here is only $+$2 and $+$1 instances and does \emph{not} reach significance, so the honest reading is \emph{on par with the best agents}, not ahead of them.
\item \textbf{Raw upstream evidence sinks the agents.}
Both agents are strong with minimal context but collapse once fed the raw changelog---Codex to 73.7\% and Claude~Code to 57.9\%, both significantly below \textsc{DepRepair} ($p{<}0.001$).
\end{itemize}

\subsection{RQ2: The Empty Patch as the Dominant Failure}\label{sec:rq2}

To analyze repair failure modes, we examine the artifact each method generates before oracle evaluation, distinguishing a produced candidate patch from an \emph{empty patch} (i.e., no code diff).
Failures rarely arise from explicit abstention; instead, the model often incorrectly concludes that the consumer code remains compatible and needs no modification.
Such false-negative localization errors manifest as empty patches despite the existence of a required repair.
Because empty patches are identifiable directly from generation logs, independent of oracle behavior, they serve as a robust proxy for localization failures.
Table~\ref{tab:abstention} summarizes each method's empty-patch frequency over the 95 benchmark instances.

\begin{table}[htbp]
\caption{Generation-side empty patches: number of instances (out of 95) on which the method emits \emph{no} patch. Raw upstream evidence drives the model to emit an empty patch far more often; \textsc{DepRepair}'s structured evidence does so least.}
\begin{center}
\begin{tabular}{lrr}
\toprule
\textbf{Method} & \textbf{Empty patches} & \textbf{Rate} \\
\midrule
\textsc{DepRepair}        & 10 & 10.5\% \\
Direct LLM (minimal)      & 21 & 22.1\% \\
Agentic Codex (minimal)   & 21 & 22.1\% \\
Direct LLM (upstream)     & 32 & 33.7\% \\
Agentic Claude~Code (minimal) & 35 & 36.8\% \\
\bottomrule
\end{tabular}
\label{tab:abstention}
\end{center}
\end{table}

Two findings stand out.
\begin{itemize}
\item \textbf{The empty patch is the dominant failure.}
Even the strongest LLM baselines decline to emit a patch on roughly a fifth of instances (21/95), and \emph{raw} upstream evidence makes this sharply worse---Direct LLM (upstream) returns an empty patch on 32/95 (33.7\%) and Agentic Claude~Code on 35/95 (36.8\%).
The largest single barrier is therefore \emph{producing any fix at all}, and unstructured evidence aggravates it.
\item \textbf{\textsc{DepRepair} attacks exactly this barrier.}
It returns an empty patch on only 10/95 instances (10.5\%), less than half the rate of Direct LLM (minimal), and---crucially---the patches it \emph{does} emit almost always pass: on GPT-5.5 every one of the 85 emitted diffs clears the executable oracle.
In other words, structured evidence converts the model's non-answers into small, applicable, oracle-passing patches.
\end{itemize}
This empty-patch gap is the largest single lever on pass rate: the raw-evidence configurations abstain on 32--35 of 95 instances, whereas \textsc{DepRepair} does so on only 10, turning non-answers into applicable patches.

\subsection{RQ3a: Evidence as a Double-Edged Sword}\label{sec:rq3a}

A surprising finding is that providing raw upstream evidence \emph{hurts} performance for both LLMs and agents.

\begin{table}[htbp]
\caption{Impact of evidence form on pass rate. Raw evidence consistently reduces performance vs.\ the no-evidence baseline.}
\begin{center}
\begin{tabular}{lccr}
\toprule
\textbf{Method} & \textbf{Minimal} & \textbf{Upstream} & \textbf{$\Delta$} \\
\midrule
Direct LLM (GPT-5.5)  & 82.1\% & 74.7\% & $-$7.4pp \\
Agentic Codex (GPT-5.5) & 87.4\% & 73.7\% & $-$13.7pp \\
Direct LLM (Opus~4.6) & 71.6\% & 61.1\% & $-$10.5pp \\
Agentic Claude~Code (Opus) & 81.1\% & 57.9\% & $-$23.2pp \\
\bottomrule
\end{tabular}
\label{tab:evidence}
\end{center}
\end{table}

Table~\ref{tab:evidence} shows that raw upstream evidence reduces pass rates by 7--23 percentage points across all methods, and two observations stand out.
\begin{itemize}
\item \textbf{Raw evidence hurts every method.}
The drops are large and directionally consistent across every method--backbone pair, and Claude~Code is the most severely affected ($-$23.2pp), possibly because its longer agent loop amplifies the misleading effect of unstructured evidence.
\item \textbf{Structured evidence reverses the effect.}
\textsc{DepRepair}'s structured evidence extraction turns the loss into a gain: pass rates \emph{increase} by 7.4pp (GPT-5.5, 82.1\%$\to$89.5\%) and 10.5pp (Opus~4.6, 71.6\%$\to$82.1\%) over the minimal baseline.
\end{itemize}

This reveals a fundamental insight: \emph{the value of upstream evidence depends entirely on how it is processed}.
Raw changelogs and migration guides contain both actionable information and irrelevant noise; LLMs and agents lack the ability to reliably separate the two, leading to worse outcomes than having no evidence at all.

\subsection{RQ3b: Component Ablation}\label{sec:rq3}

Table~\ref{tab:ablation} presents the ablation results for both backbones.

\begin{table}[htbp]
\caption{Ablation study: contribution of each \textsc{DepRepair} component.}
\begin{center}
\begin{tabular}{lccccrc}
\toprule
& \multicolumn{3}{c}{\textbf{Components}} & & \multicolumn{2}{c}{\textbf{Result}} \\
\cmidrule{2-4} \cmidrule{6-7}
\textbf{Configuration} & Rules & Loc. & SubCat & & Pass & Rate \\
\midrule
\multicolumn{7}{l}{\emph{Claude Opus~4.6 backbone}} \\
Full \textsc{DepRepair}    & \checkmark & \checkmark & \checkmark & & \textbf{78} & \textbf{82.1\%} \\
$-$Subcategory             & \checkmark & \checkmark & ---        & & 58 & 61.1\% \\
$-$Evidence                & ---        & ---        & \checkmark & & 58 & 61.1\% \\
Minimal (no components)    & ---        & ---        & ---        & & 68 & 71.6\% \\
\midrule
\multicolumn{7}{l}{\emph{GPT-5.5 backbone}} \\
Full \textsc{DepRepair}    & \checkmark & \checkmark & \checkmark & & \textbf{85} & \textbf{89.5\%} \\
$-$Subcategory             & \checkmark & \checkmark & ---        & & 82 & 86.3\% \\
$-$Evidence                & ---        & ---        & \checkmark & & 80 & 84.2\% \\
Minimal (no components)    & ---        & ---        & ---        & & 78 & 82.1\% \\
\bottomrule
\end{tabular}
\label{tab:ablation}
\end{center}
\end{table}

The two backbones expose complementary effects, but converge on one message: evidence and guidance only help when supplied together.
\begin{itemize}
\item \textbf{On Opus, a half-pipeline is worse than none.}
Removing either subcategory instructions or structured evidence drops the pass rate by 21.0pp (82.1\%$\to$61.1\%); both contrasts against the full pipeline are large enough to be highly significant under McNemar ($p{<}0.001$).
Strikingly, both ablated configurations (61.1\%) land \emph{below} the minimal baseline with no components at all (71.6\%).
This mirrors the evidence pattern in Section~\ref{sec:rq3a}: structured rules \emph{without} category-specific guidance act as noise, just as raw evidence does, because the model cannot tell which of the supplied rules apply.
\item \textbf{The contribution is backbone-dependent.}
On the weaker Opus backbone, the full pipeline adds 10.5 percentage points over the minimal configuration. In contrast, the stronger GPT-5.5 backbone already achieves 82.1\% with the minimal prompt and therefore benefits less from the additional components. Removing either component still retains 84.2--86.3\%, remaining above the minimal baseline.
\end{itemize}
The consistent takeaway across backbones is that \emph{evidence and guidance must work together}: supplying one without the other is at best neutral and at worst actively harmful.

\subsection{RQ4: Per-Subcategory Analysis}\label{sec:subcategory}

Table~\ref{tab:subcategory} breaks down \textsc{DepRepair}'s performance by breaking-change subcategory.

\begin{table}[htbp]
\caption{Pass rates by breaking-change subcategory (best methods). Counts in parentheses are the number of instances per subcategory.}
\begin{center}
\begin{tabular}{lccccc}
\toprule
\textbf{Method} & \textbf{CAM} & \textbf{DR} & \textbf{IM} & \textbf{PS} & \textbf{All} \\
& (25) & (27) & (23) & (20) & (95) \\
\midrule
\textsc{DepRepair} (GPT-5.5) & \textbf{96\%} & \textbf{93\%} & \textbf{100\%} & 65\% & \textbf{89.5\%} \\
Agentic Codex (GPT)  & 92\% & 89\% & \textbf{100\%} & 65\% & 87.4\% \\
Direct LLM (GPT-5.5) & 88\% & 81\% & \textbf{100\%} & 55\% & 82.1\% \\
\textsc{DepRepair} (Opus)    & 80\% & 78\% & \textbf{100\%} & \textbf{70\%} & \textbf{82.1\%} \\
Agentic Claude~Code (Opus)& 84\% & 81\% & \textbf{100\%} & 55\% & 81.1\% \\
Direct LLM (Opus)    & 64\% & 63\% & \textbf{100\%} & 60\% & 71.6\% \\
\bottomrule
\end{tabular}
\label{tab:subcategory}
\end{center}
\end{table}

Table~\ref{tab:subcategory} exposes a clear difficulty gradient across subcategories.
\begin{itemize}
\item \textbf{Import migration is saturated.}
Every method in Table~\ref{tab:subcategory} achieves 100\% on import migration, so it offers no room to separate methods.
\item \textbf{\textsc{DepRepair}'s gains concentrate on CAM and DR.}
\textsc{DepRepair}'s advantage lands on compound API migration (CAM) and direct rename (DR), where structured evidence most directly helps: on the Opus backbone it lifts CAM from 64\% to 80\% ($+$16pp) and DR from 63\% to 78\% ($+$15pp) over the minimal baseline, and on GPT-5.5 it lifts CAM from 88\% to 96\% ($+$8pp) and DR from 81\% to 93\% ($+$12pp).
\item \textbf{Paradigm shift is the residual gap.}
Paradigm shift (PS) is now the hardest subcategory: even \textsc{DepRepair} reaches only 65\% (GPT-5.5) and 70\% (Opus), and the minimal baselines fall to 55--60\%.
It is also lifted---$+$10pp on both backbones (GPT 55\%$\to$65\%, Opus 60\%$\to$70\%)---but remains the residual gap, as it demands structural refactoring beyond the rule-level edits that structured evidence most readily supplies.
\end{itemize}
Broken down by ecosystem, \textsc{DepRepair} (GPT-5.5) solves all 38 Maven instances and 14 of 15 Cargo instances, with npm (26/33) and PyPI (7/9) trailing; the PyPI pocket---once a complete blind spot---is now largely repairable, leaving paradigm-shift instances as the clearest target for future work.

\subsection{Case Studies}\label{sec:cases}

We close with three real \textsc{DepRepair} patches (Figs.~\ref{fig:case-cam}--\ref{fig:case-ps}) that make the RQ4 difficulty gradient concrete, spanning compound API migration (CAM), direct rename (DR), and paradigm shift (PS) across three ecosystems---Maven, PyPI, and npm.
Each shows how, once the upstream change is distilled into migration rules and localized to the affected sites, a single call emits a compact edit whose \emph{shape} matches the kind of migration.

\begin{figure}[t]
\begin{tcolorbox}[colback=white,colframe=black,boxrule=0.5pt,arc=1pt,
  left=3pt,right=3pt,top=2pt,bottom=2pt,
  title=\footnotesize\textbf{exec-068\,: JUnit 4$\to$5 (Maven, CAM)}]
\begin{lstlisting}[style=patch]
-import org.junit.Assert;
-import org.junit.Before;
-import org.junit.runner.RunWith;
+import org.junit.jupiter.api.Assertions;
+import org.junit.jupiter.api.BeforeEach;
+import org.junit.jupiter.api.extension.ExtendWith;
-import org.mockito.junit.MockitoJUnitRunner;
+import org.mockito.junit.jupiter.MockitoExtension;
@@
-@RunWith(MockitoJUnitRunner.class)
+@ExtendWith(MockitoExtension.class)
 public class AdapterFactoryTest {
-    @Before
+    @BeforeEach
     public void setup() { ... }
-        Assert.assertTrue(factory.isModelClass(...));
+        Assertions.assertTrue(factory.isModelClass(...));
\end{lstlisting}
\end{tcolorbox}
\caption{CAM. \textsc{DepRepair} applies the JUnit~4$\to$5 namespace move, the Mockito runner-to-extension switch, and the annotation/assertion renames together in one coordinated patch.}
\label{fig:case-cam}
\end{figure}

\begin{figure}[t]
\begin{tcolorbox}[colback=white,colframe=black,boxrule=0.5pt,arc=1pt,
  left=3pt,right=3pt,top=2pt,bottom=2pt,
  title=\footnotesize\textbf{exec-180\,: Pydantic 1$\to$2 (PyPI, DR)}]
\begin{lstlisting}[style=patch]
-from pydantic import BaseModel, Field, validator
+from pydantic import BaseModel, ConfigDict, Field, validator
@@
-        ... for key, value in self.dict().items() ...
+        ... for key, value in self.model_dump().items() ...
@@
-        return {a for a in cls.__dict__["__fields__"].keys() ...}
+        return {a for a in cls.model_fields.keys() ...}
@@
-        self.__fields__[field].alias for field in self.__fields_set__
+        self.model_fields[field].alias for field in self.model_fields_set
\end{lstlisting}
\end{tcolorbox}
\caption{DR. The Pydantic~v2 renames are pure symbol substitutions; \textsc{DepRepair} sweeps every occurrence and adds the \texttt{ConfigDict} import that the new names require.}
\label{fig:case-dr}
\end{figure}

\begin{figure}[t]
\begin{tcolorbox}[colback=white,colframe=black,boxrule=0.5pt,arc=1pt,
  left=3pt,right=3pt,top=2pt,bottom=2pt,
  title=\footnotesize\textbf{exec-841\,: Vue 2$\to$3 (npm, PS)}]
\begin{lstlisting}[style=patch]
 import {
   defineComponent,
-  getCurrentInstance,
   h,
+  nextTick,
   ref,
 } from 'vue';
@@
       expose(publicApi);
-      const instance = getCurrentInstance();
-      instance?.proxy?.$nextTick(() => {
+      nextTick(() => {
         for (const key in innerRef.value) { ... }
\end{lstlisting}
\end{tcolorbox}
\caption{PS. The Vue~2 instance-proxy idiom has no symbol-level equivalent; \textsc{DepRepair} restructures it into the standalone \texttt{nextTick} import and call---a pattern rewrite, not a rename.}
\label{fig:case-ps}
\end{figure}

\begin{itemize}
\item \textbf{exec-068 (junit, CAM).}
The structured rules tie the JUnit~4$\to$5 namespace move, the Mockito runner-to-extension switch (\texttt{@RunWith(MockitoJUnitRunner)}$\to$ \\ \texttt{@ExtendWith(MockitoExtension)}), and the \texttt{@Before}$\to$\texttt{@BeforeEach} / \texttt{Assert}$\to$\texttt{Assertions} renames into one change; \textsc{DepRepair} applies all of them together, as a partial CAM that omits any one would leave the test uncompilable.
\item \textbf{exec-180 (pydantic, DR).}
The Pydantic~v2 renames are pure symbol substitutions---\texttt{.dict()}$\to$\texttt{.model\_dump()}, \texttt{\_\_fields\_\_}$\to$\texttt{model\_fields}, \texttt{\_\_fields\_set\_\_}$\to$\texttt{model\_fields\_set}---which \textsc{DepRepair} sweeps exhaustively while pulling in the \texttt{ConfigDict} import the new API needs.
\item \textbf{exec-841 (vue, PS).}
The Vue~2 idiom \texttt{getCurrentInstance().proxy.\$nextTick(...)} has no one-to-one Vue~3 replacement; \textsc{DepRepair} rewrites it into the standalone \texttt{nextTick(...)} import and call---the kind of structural pattern change (not renaming) that keeps PS the residual gap (RQ4).
\end{itemize}
Read together, the three trace one capability at rising difficulty: a coordinated multi-API rewrite for CAM, a mechanical symbol sweep for DR, and a structural replacement for PS---each landing precisely on the migrated API surface that the structured evidence identified.

\section{Discussion}\label{sec:discussion}

\textbf{Why structured evidence wins.}
The dominant failure is the \emph{empty patch}: facing a raw changelog, the model must decide \emph{what} changed, \emph{where} it matters, and \emph{how} to edit at once, and often does none---returning no patch (32 of 95 for Direct LLM (upstream) and 35 for Agentic Claude~Code, versus 10 for \textsc{DepRepair}).
\textsc{DepRepair} factors this decision---evidence extraction answers \emph{what}, localization \emph{where}, and subcategory guidance \emph{how}---converting non-answers into small, oracle-passing patches.
The value of cross-repository context thus lies in its \emph{processing}, not its mere \emph{provision} (RQ3): raw text is counterproductive, whereas distilled rules reliably help.
We do not overclaim: in a single call \textsc{DepRepair} \emph{significantly} beats the same-backbone Direct LLM (minimal) and every raw-evidence configuration, while only \emph{matching} the strongest cross-architecture agent on each backbone.

\textbf{Threats to validity.}
We organize threats along the four standard dimensions.
\begin{itemize}
\item \emph{Statistical conclusion validity.} All methods run on the same 95 instances, so we use the \emph{paired} McNemar test (exact binomial form when discordant pairs are few) and Fisher's exact test as a conservative check. At $n{=}95$ power is limited, so we separate \emph{ranking} claims (\textsc{DepRepair} is first on each backbone) from \emph{significance} claims, stating only that it significantly beats the same-backbone Direct LLM (minimal) and all raw-evidence configurations while \emph{matching} the strongest agent.
\item \emph{Construct validity.} The oracle counts an instance as repaired iff the breakage-related tests pass, which cannot certify correctness beyond test coverage; we mitigate this with the developer's own test suite and executable verdicts rather than file-level equivalence. Subcategory labels are author-assigned and released for scrutiny.
\item \emph{Internal validity.} Outputs are re-applied and re-executed in freshly rebuilt containers, so the oracle is deterministic; generation uses temperature~0 with no retries, and timeouts or errors count as failures.
\item \emph{External validity.} Results may not generalize beyond the studied packages and ecosystems; we mitigate single-model bias with two very different backbones (GPT-5.5 and Claude~Opus~4.6) that show consistent trends.
\end{itemize}

\textbf{Limitations.}
\begin{itemize}
\item \emph{Data volume.} \textsc{DepBench} currently comprises 95 instances, which limits statistical power; we are expanding it toward 200 instances to sharpen significance and broaden coverage.
\item \emph{Data distribution.} Instances are unevenly spread across ecosystems and subcategories (e.g., Maven~38 vs.\ PyPI~9; paradigm shift only~20), so some per-cell rates rest on small samples; a more balanced distribution would make subcategory comparisons more robust.
\end{itemize}

\textbf{Generalizability.}
While evaluated on four ecosystems, \textsc{DepRepair}'s approach is ecosystem-agnostic: the upstream evidence extraction, localization, and subcategory-aware prompting stages apply to any language.
The consistent improvement across two very different LLM backbones (GPT-5.5 and Claude~Opus~4.6) suggests that \textsc{DepRepair}'s benefits are not model-specific.

\section{Conclusion}\label{sec:conclusion}

We presented \textsc{DepRepair}, an LLM-based approach to source-code repair for dependency breaking changes.
By combining upstream evidence extraction, consumer-side usage localization, and subcategory-aware prompt construction, \textsc{DepRepair} attains the highest executable pass rate on each backbone---89.5\% (GPT-5.5) and 82.1\% (Claude~Opus~4.6)---on a benchmark of 95 real-world instances, \emph{significantly} beating the same-backbone Direct LLM (minimal) baseline on both backbones and ranking ahead of state-of-the-art coding agents.
A generation-side analysis shows that the \emph{empty patch}---the model emits no usable patch---is the dominant failure mode: raw upstream evidence drives empty patches on up to a third of instances, whereas \textsc{DepRepair} emits one on only 10 of 95 and nearly every patch it does emit passes, identifying the empty patch as the largest single barrier that structured evidence relieves.
Our analysis further reveals that raw upstream evidence is a double-edged sword---hurting both LLMs (up to $-$10.5pp) and agents (up to $-$23.2pp)---while structured evidence extraction is essential for effective repair.
These findings highlight the importance of cross-repository context \emph{processing}, not merely \emph{provision}, in LLM-based program repair.

\section*{Acknowledgment}
Generative AI tools were used to assist with drafting and editing portions of this manuscript.
All technical content, experimental results, and conclusions were produced and verified by the authors.


\bibliographystyle{IEEEtran}
\bibliography{references1}

\end{document}